\documentclass[a4paper,twocolumn,prl,superscriptaddress]{revtex4}

\usepackage{bm,amsmath}
\usepackage{graphicx}
\usepackage{xcolor}
\usepackage{hyperref}
\hypersetup{
    colorlinks,
    linkcolor={red!90!black},
    citecolor={blue!90!red},
    urlcolor={blue!100!black}
}

\usepackage{wasysym}
\usepackage{amssymb}
\usepackage{stackengine}
\usepackage{scalerel}
\newcommand\openbigstar[1][0.7]{%
  \scalerel*{%
    \stackinset{c}{-.125pt}{c}{}{\scalebox{#1}{\color{white}{$\bigstar$}}}{%
      $\bigstar$}%
  }{\bigstar}
}

\newcommand{\f}{\frac}

\renewcommand{\l}{\ell^*}
\newcommand{\D}{\Delta}

\def\[{\left[}
\def\]{\right]}
\def\({\left(}
\def\){\right)}
\def\be{\begin{equation}}
\def\ee{\end{equation}}
\def\bea{\begin{eqnarray}}
\def\eea{\end{eqnarray}}

\begin{document}

\title{Activity controls fragility: A Random First Order Transition Theory for an active glass}

\author{Saroj Kumar Nandi}
\affiliation{Department of Chemical and Biological Physics, Weizmann Institute of Science, Rehovot 7610001, Israel}
\email{saroj.nandi@weizmann.ac.il}

\author{Rituparno Mandal}
\affiliation{Department of Physics, Indian Institute of Science, Bangalore 560012, India}

\author{Pranab Jyoti Bhuyan}
\affiliation{Department of Physics, Indian Institute of Science, Bangalore 560012, India}

\author{Chandan Dasgupta}
\affiliation{Department of Physics, Indian Institute of Science, Bangalore 560012, India}
\affiliation{International Centre for Theoretical Sciences, Bangalore 560089}

\author{Madan Rao}
\affiliation{Simons Centre for the Study of Living Machines, National Centre for Biological Sciences (TIFR), Bangalore 560065, India}

\author{Nir S. Gov}
\affiliation{Department of Chemical and Biological Physics, Weizmann Institute of Science, Rehovot 7610001, Israel}

\begin{abstract}
How does nonequilibrium activity modify the approach to a glass? This is an important question, since many experiments reveal the near-glassy nature of the cell interior, remodelled by activity.
However, different simulations of dense assemblies of active particles, parametrised by a self-propulsion force, $f_0$, and persistence time, $\tau_p$, 
appear to make contradictory predictions about the influence of activity on characteristic features of glass, such as fragility. This calls for a broad conceptual framework to understand active glasses; here we extend the Random First-Order Transition (RFOT) theory to a dense assembly of self-propelled particles. We compute the active contribution to the configurational entropy using an effective medium approach - that of a single particle in a caging-potential. This simple active extension of RFOT provides excellent quantitative fits to existing simulation results. We find that whereas $f_0$ always inhibits glassiness, the effect of $\tau_p$ is more subtle and depends on the microscopic details of activity. In doing so, the theory automatically resolves the apparent contradiction between the simulation models. The theory also makes several
 testable predictions, which we verify by both existing and new simulation data, and should be viewed as a step towards a more rigorous analytical treatment of active glass.
\end{abstract}
\maketitle

Active systems, consisting of particles that convert energy supplied to it into mechanical work, are a fascinating class of driven nonequilibrium systems~\cite{sriramreview,sriramrmp}. 
The range of systems that fall under this ambit, include {\it living systems}, from cells and their motor-cytoskeletal extracts to collections of cells constituting tissues, which utilise chemical 
energy to perform biological function~\cite{jacques2015}, and {\it synthetic systems}, such as magnetic colloidal beads~\cite{dreyfus2005}, light-activated colloidal swimmers~\cite{palacci2013}, 
and vertically vibrated grains~\cite{nitin2014,dauchot2005}. So far, theoretical and experimental studies have focussed on dilute collections of active particles, however recent experiments 
on the collective movement of confluent cells in tissues have studied the effects of activity in these dense regimes, where the dynamics is slow, multi-particle correlations are significant  and the system 
approaches a collective jammed or glassy state~\cite{zhou2009,angelini2011,garcia2015,sadati2014,parry2014,gravish2015}.

Motivated by these, there have been a number of simulation studies~\cite{kranz2010,berthier2013,ni2013,peruani2013,mandal2016,berthier2014,bi2015,ni2014,levis2015,ginot2015,szamel2016,flenner2016,behchinger2016,redner2016} 
of dense assemblies of stochastically driven particles, parametrised by a self-propulsion force of magnitude $f_0$ and an orientational persistence time $\tau_p$. Each of these models
characterises the statistics of activity in a slightly different way, however their predictions on the variation of signature glassy behaviour such as fragility, 
are divergent. Unlike in the case of dilute collections of active particles, there is no analytically calculable framework to understand the dynamics of active particles in the dense limit. 
There have been attempts to extend the usual Mode-Coupling theory (MCT) to include active self-propulsion~\cite{kranz2010,berthier2013,szamel2016,noneqMCT}, however such 
theories are not applicable for the glassy dynamics at high densities~\cite{giulioreview,manoj2015}.

\begin{figure}
\includegraphics[width=7.4cm]{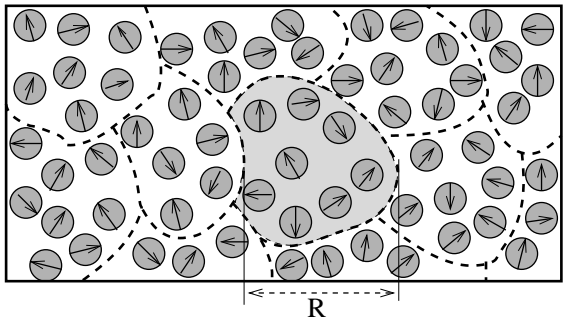}
\caption{Schematic picture of the mosaic concept of RFOT. The system consists of different regions of unique configurations of a typical length scale $\l$.
Activity affects the system when $\l$ decreases with $f_0$ and either increases or decreases with $\tau_p$ depending on the nature of activity. The dotted lines are schematic representation for a cluster of particles in the same state. Arrows on the particles indicate the instantaneous directions along which the self-propelled particles apply their motile force $f_0$, for a time scale of $\tau_p$.}
\label{mosaicsc}
\end{figure}

The Random First-Order Transition (RFOT) theory~\cite{kirkpatrick1987,kirkpatrick1989,lubchenko2007,giuliobook,kirkthiruRMP} has been remarkably successful~\cite{xia2000,wolynes2012,wolynes2009} 
in describing a glassy system both above and below the regime where activated processes dominate the dynamics.  RFOT theory generalises the theory of first-order crystallisation transition to that 
of freezing to a disordered structure~\cite{kirkthiruRMP,lubchenko2007}, describing the system in terms of a {\it mosaic of local aperiodic domains} (Fig.\,\ref{mosaicsc}) and a mismatch energy at the interface between domains \cite{xia2000,wolynes2009,wolynes2012}. Including activity within this basic RFOT picture presents a major challenge (see~\cite{supmat}). Here we propose an extension of  the 
RFOT theory to an active system, that should be viewed as a first step towards a more rigorous analytical treatment of active glass. Even so, the theory presented here
makes several testable predictions, regarding how the active parameters affects glassy behaviour - whereas $f_0$ always inhibits glassiness, the effect of $\tau_p$ is more subtle 
 and depends on the microscopic details of activity. This not only helps reconcile the apparently divergent results of the simulation models discussed above, but also makes predictions which we verify
with new numerical simulations ({\it Methods}).

\subsection*{Active RFOT : General Formalism}
The RFOT theory for passive glasses describes the activated reconfigurations of such mosaics through a mechanism similar to nucleation~\cite{kirkpatrick1987,kirkpatrick1989,xia2000,lubchenko2007,lubchenko2014}. Consider a domain of volume $v$ and length scale $R\sim v^{1/d}$, in $d$-dimension, 
as shown by the shaded region in Fig.\,\ref{mosaicsc}. The RFOT theory posits that the driving force for reconfiguration of the mosaics is the configurational entropy density,
$s_c(\Phi, T)$, with potential energy $\Phi$ at temperature $T$. We can write the free energy gain of the system due to such reconfiguration as
\begin{equation} \label{arg1}
 \Delta F(R)=-\Omega_d\,R^d\,T\,s_c(\Phi, T)+S_d\,R^\theta\,\bar{\gamma},
\end{equation}
where $\Omega_d$ is the volume and $S_d$ the surface of a region with unit radius in dimension $d$ and $\bar{\gamma}$ is the surface mismatch energy. 
The driving force responsible for the reconfiguration of a domain of size $R$ is obtained by differentiating Eq.\,(\ref{arg1}), with respect to $R$. The critical 
nucleation size $\ell^*$ is determined by force-balance.
If the cavity size $R > \l$, then the volume contribution drives the reconfiguration of the mosaic.
On the other hand, if $R < \l$,  the surface contribution dominates and the state within the cavity remains the same. Thus, the system consists of mosaics of correlated domains with a typical length scale $\ell^*$. Note that this length-scale may be different from dynamic length-scales associated with domains of correlated motion~\cite{garcia2015,annrev,mandal2017}.

In extending the conventional RFOT theory to active systems
that are manifestly out of equilibrium, we must first find a physically reasonable definition of the configurational entropy $s_c$ for such systems. The configurational entropy of passive thermal systems is usually defined in terms of the multiplicity of the local minima of the potential energy (``inherent structures'') whose basins of interaction are visited by the system at a particular temperature $T$. We continue to use the same definition of the configurational entropy for active systems. Particle configurations in a steady state of an active systems can be used as starting points of a minimisation procedure to locate a set of inherent structures characteristic to the steady state being considered. The multiplicity of inherent structures obtained this way would be different from that of the inherent structures obtained for an identical system without activity at the same temperature. Therefore, the temperature dependence of the configurational  entropy of an active system would be different from that for its passive counterpart.  


To proceed further we need to evaluate the active corrections to the configurational entropy. Since the typical persistence time
of the active force is larger than the time scale of thermal fluctuations, the effects of activity, to leading order, can 
be accommodated in a renormalised potential energy, $\tilde \Phi=\Phi+\delta \Phi$, whose form depends on the precise model of activity~\cite{fodor2016,marconi2016}. 
We assume that $\delta \Phi = -\sum_i \langle {\bf f}^a_i \cdot {\bf x}_i \rangle$ where ${\bf{x}}_i$ is the position vector of particle $i$, 
${\bf f}^a_i$ is the active force and $\langle \cdots \rangle$ denotes an average over a steady-state distribution of the particle positions \cite{supmat}.
Expanding the active configurational entropy about its value for the passive system, we get
\bea
s_c[{\tilde \Phi}] &=&s_c[\Phi] +  \int  \delta \Phi(\{x_i\}) \frac{\delta s_c[\Phi]}{\delta \Phi(\{x_i\})}\bigg|_{{\bf f}^{a}=0, T}\,\prod_i dx_i \nonumber \\ 
& &  + \ldots 
\label{eq:confentropy}
\eea
 where 
the leading correction involves the functional derivative of the configurational entropy evaluated at zero active force ${\bf f}_i^a$. 
We write this correction term as $\kappa_a \langle {\bf f}^a\cdot {\bf x} \rangle$ where we have introduced an
{\it active fragility parameter}, $\kappa_a$ that quantifies
the sensitivity of the configurational entropy to changes in active force. In addition, to arrive at an active extension of RFOT, we must also evaluate the active surface contribution. Here we simply take it to be
$\bar{\gamma}(f_0,\tau_p,T)$ (Eq. \ref{arg1}).

Since the Kauzmann temperature $T_K$ is defined as the temperature where the configurational entropy  $s_c(\Phi,T)$ of the passive system vanishes~\cite{kauzmann1948}, we take, 
following the prescription in~\cite{lubchenko2007,kirkthiruRMP}, $s_c(\Phi,T)=\D C_p(T-T_K)/T_K$ close to but above $T_K$, where $\Delta C_p$ is the jump in specific heat from the
 liquid to the crystalline state~\cite{kauzmann1948}.  Further, we assume, as in the passive glass~\cite{wolynesbook},  that the temperature-dependence of $\bar{\gamma}$ is linear:  
 $\bar{\gamma} = \kappa(f_0,\tau_p)\,T$, where $\kappa$ is a function of the active parameters alone. 
As in the passive glass, the detailed $T$-dependence of $\bar{\gamma}$ does not change the qualitative results because the change in configurational entropy plays the major 
role in governing the dynamics~\cite{wolynesbook}.
Defining $D\equiv c_d\kappa T_K/\D C_p$, where $c_d=\theta S_d/d\Omega_d$,
we obtain, close to $T_K$,
\begin{equation}\label{lstareq1}
 \l=\left[\f{{D}}{(T-T_K)+\frac{T_K {\kappa_a} \langle   {\bf f}^a \cdot {\bf x}\rangle}{\Delta C_p}}\right]^{1/(d-\theta)}.
\end{equation}
This expression is similar in spirit to what Wang and Wolynes proposed for active network materials~\cite{wang2013}. Note that while in principle, $D$ depends on the active parameters, $f_0,\tau_p$, for simplicity we will consider it to be independent of the activity.

The relaxation dynamics of the system is characterised by the reconfiguration of a domain of length scale $\l$. The typical potential energy barrier height for the relaxation of the system is $\D(\l)\sim \D_0 {\l}^\psi$,
where $\D_0$ is a $T$-dependent energy scale and $\psi$, an exponent. Then the relaxation time $\tau$ of the system is $\tau \sim \tau_0\exp\left(\f{\D_0{\l}^\psi}{T}\right)$,
where $\tau_0$ is a microscopic time scale determined by the inter-particle interactions.
If, following \cite{kirkthiruRMP,kirkpatrick1989}
we take $\psi=\theta=d/2$, and assuming $\D_0=\bar{\gamma}$, we 
obtain, close to $T_K$,
\begin{equation}
\label{taueq1}
 \ln\left(\f{\tau}{\tau_0}\right)=
 \f{{E}}{(T-T_K)+\f{T_K{\kappa_a} \langle  {\bf f}^a \cdot {\bf x}\rangle}{\D C_p}}
\end{equation}
where $E=\kappa D=c_d\kappa^2T_K/\D C_p$. The alternative suggestion~\cite{annrev}, $\theta = d-1$, $\psi=1$, would change the expression for $\l$, but lead to the same expression for $\tau$. Therefore,
within this active extension of the RFOT, all one needs is the activity correction to the configurational entropy ${\kappa_a} \langle  {\bf f}^a \cdot {\bf x}\rangle$. The sign of this correction is important because it tells us whether the presence of activity increases or decreases the configurational entropy. We have carried out test simulations (see~\cite{supmat} for details) to check how the average energy of inherent structures in the model of active glass considered in~\cite{mandal2016} depends on the strength of activity $f_0$ (see Fig. S1). We find that the average energy of inherent structures increases as $f_0$ is increased at a fixed temperature. This observation implies that the configurational entropy also increases with $f_0$, i.e. the sign of the activity correction is positive.
This conclusion is consistent with the results of recent work by Preisler and Dijkstra (Fig.\,S2)~\cite{supmat,preisler2016}. A positive entropy correction  implies that the temperature at which $\tau$ diverges for an active system is lower than $T_K$, i.e. the presence of activity decreases the glass transition temperature.

\subsection*{Active RFOT : Application to models of self-propelled particles}
We will now explicitly evaluate the active correction in Eq.\,\ref{eq:confentropy},
for a system of self-propelled particles~\cite{sriramreview,sriramrmp,ni2013,mandal2016},
described by the following overdamped dynamics
  \be
 \partial_t {\bf x}_i = - \mu \nabla_i \Phi_i + {\bf f}^{a}_i(t)+ {\bf f}_i(t)
  \label{eq:active}
 \ee
  where ${\bf x}_i$ is the position of the $i$-th particle, $\mu$ is the particle mobility,
  $\Phi_i \equiv \sum_{j\neq i} \phi({\bf x}_i, {\bf x}_j)$ is the many-body potential experienced by particle $i$,
  and ${\bf f}^{a}_i(t)$ and ${\bf f}_i(t)$ are the active and thermal noises, respectively. 
This correction, evaluated using an approximate effective medium approach, depends on the precise statistical nature of activity~\cite{supmat}. 
As detailed in~\cite{supmat}, our effective medium approach consists of replacing the many-particle dynamics (Eq.\,\ref{eq:active}), by the dynamics of a single self-propelled particle caged by the other particles,
whose effect is represented as an effective confining harmonic potential of strength $k$~\cite{benisaac2015,noneqMCT,manoj2017}.
 We then calculate $\langle  {\bf f}^a \cdot {\bf x}\rangle$ where the angular bracket is an average over the  steady state probability distribution of the caged active particle.
We consider two different models of active forces that have been widely used in the simulation literature and show that they give fundamentally different behaviour as a function of $\tau_p$, although the dependence on $f_0$ is the same for both models.  These opposite effects of the propulsion force and the persistence time on the glass transition in active glasses, that were reported
in simulations~\cite{mandal2016,flenner2016} was a
puzzle, and finds a natural resolution in our theoretical framework.\\

\noindent
\text{{\bf Model 1:}} The active noise 
${\bf f}^a_i$ with fixed amplitude $f_0$, has zero mean, and a shot-noise temporal correlation
\begin{equation}\label{model1}
 \langle {\bf f}^a_i(0) \cdot {\bf f}^a_i(t)\rangle = f_0^2\exp\left[-t/\tau_p\right]\,.
\end{equation}
This is the realisation of the active noise considered in the simulation studies of~\cite{mandal2016}.
For this model of active forces,  the active RFOT calculation detailed in~\cite{supmat} gives for the correction to the configurational entropy appearing in Eqs.\,(\ref{lstareq1})\,and\,(\ref{taueq1}),
\begin{equation}
 \f{T_K \kappa_a \langle  {\bf f}^a \cdot {\bf x}\rangle}{\Delta C_p}
 = \f{Hf_0^2\tau{_p}}{(1+G\tau_p)}.\label{faeq2}
\end{equation}
where $H = \f{T_K \kappa_a}{\gamma\,\Delta C_p}$ and $G=k/\gamma$. For $\theta = \psi = d/2$, we obtain for Eqs.\,(\ref{lstareq1})\,and\,(\ref{taueq1}),
\begin{align}
& \l=\left[\f{D}{(T-T_K)+\f{Hf_0^2\tau_p}{1+G\tau_p}}\right]^{2/d} \label{length}\\
& \ln\left(\f{\tau}{\tau_0}\right)=\f{E}{(T-T_K)+\f{Hf_0^2\tau_p}{1+G\tau_p}}. \label{relation_time}
\end{align}
Eq.\,(\ref{length}) shows that both $f_0$ and $\tau_p$ decrease the effective Kauzmann temperature $T_K^{eff}=T_K-Hf_0^2\tau_p/(1+G\tau_p)$, defined as the temperature when $\l$ and $\tau$ diverge.
In other words, as we increase either $f_0$ or $\tau_p$, the system shows glassy behaviour at lower $T$ compared to the corresponding passive system, i.e., both $f_0$ and $\tau_p$ promote 
{\it fluidisation} (or suppress glassiness) for small $\tau_p$. This prediction with respect to $f_0$ is consistent with many simulation studies~\cite{berthier2013,berthier2014,mandal2016,ni2013,mandal2017}.\\

\noindent
\text{{\bf Model 2:}} The active noise ${\bf f}^a_i$ with fixed single-particle effective temperature $T_{eff}^{sp}$ is an Ornstein-Uhlenbeck 
process~\cite{uhlenbeck1930}, with correlations,
\begin{equation}\label{model2}
\langle {\bf f}_a(0) \cdot {\bf f}_a(t)\rangle = (T_{eff}^{sp}/\tau_p)\exp[-t/\tau_p].
\end{equation}
This is the realisation of the active noise considered in the simulation studies of~\cite{flenner2016}.
For this model of active forces,  the active RFOT calculation detailed in~\cite{supmat} gives
\begin{equation}
 \f{T_K \kappa_a \langle  {\bf f}^a \cdot {\bf x}\rangle}{\Delta C_p} = {HT_{eff}^{sp}}/{(1+G\tau_p)}.
\end{equation}
where again $H = \f{T_K \kappa_a}{\gamma\,\Delta C_p}$ and $G=k/\gamma$. This implies, that for $\theta = \psi = d/2$, we get
\begin{align}
&\l =\left[\f{D}{(T-T_K)+\f{HT_{eff}^{sp}}{1+G\tau_p}}\right]^{2/d}  \label{lmodel2}\\
&\ln\left[\f{\tau}{\tau_0}\right] =\f{E}{(T-T_K)+{HT_{eff}^{sp}}/(1+G\tau_p)} \label{taumodel2}
\end{align}
While the effect of $T_{eff}^{sp}$ remains same as that of $f_0^2$ in Model 1, the effect of changing $\tau_p$ is entirely different.  At a fixed $T_{eff}^{sp}$, an increase $\tau_p$ now makes the system show glassy behaviour at higher $T$, i.e., $\tau_p$ {\it promotes glassiness}. This prediction with respect to $\tau_p$ is consistent with another set of simulations~\cite{flenner2016}.

This apparent contradiction reported in simulations, that activity may either promote {\it fluidisation} or {\it glassiness}, finds a natural resolution within our active RFOT framework.
The trends strongly depend on the microscopic details of activity. We now study other predictions that emerge from our active RFOT theory and check to see if they are borne out by 
our simulations of an active glass model.


\begin{figure}
\includegraphics[width=8.6cm]{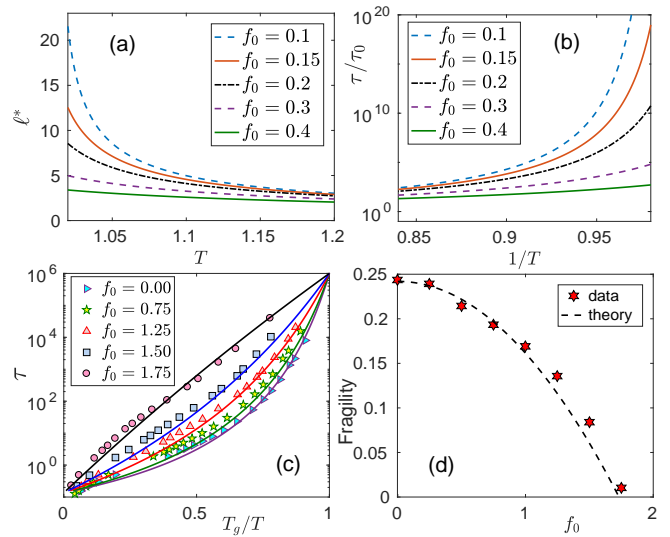}
\caption{Results when the self-propulsion force $f_0$ is the control parameter: (a) Behaviour of typical length scale $\l$ of the mosaics of an active system as a function of temperature $T$ at constant persistence time $\tau_p=0.02$, according to Eq.\,(\ref{length}). At all $T$, $\l$ decreases with increasing self-propulsion, $f_0$.
(b) Relaxation time $\tau$ of the active system as a function of $1/T$ with $\tau_p=0.02$, according to Eq.\,(\ref{relation_time}). As $f_0$ increases, growth of $\tau/\tau_0$ becomes slower and the system relaxes faster at a certain $T$. (c) Angell plot for $\tau$ as a function of $T_g/T$. The data are from~\cite{mandal2016} and the lines are our theoretical calculations using Eq.\,(\ref{relt_comp}) (see~\cite{supmat}). 
As activity increases, the behaviour of $\tau$ becomes closer to Arrhenius law, making the system a stronger glass former. (d) Behaviour of fragility as a function of activity. Data taken from MD
simulations~\cite{mandal2016} and lines are from our theoretical calculation, Eqs.\,(\ref{relt_comp})\,and\,(\ref{fragility_sim}).
 }
 \label{reltime}
\end{figure}

\subsection*{Active RFOT confronts simulations - tuning $f_0$}
We now perform molecular dynamics simulations of a glass former driven by active propulsion forces in 3-dimensions ({\it Methods}) and make a quantitative comparison of the results obtained with our theoretical predictions.
We first tune $f_0$ at fixed $\tau_p$, in this case, both Models 1 and 2 show similar behaviour.
We plot $\l$ and $\tau$ as a function of $T$ in Figs.\,\ref{reltime}(a) and (b) respectively~\cite{supmat}.
We see that at a given $T$, $\l$ as well as $\tau$ decrease as we increase $f_0$. 
Next, we calculate the modification to the glass transition temperature $T_g$ by the activity. We define $T_g$ as the temperature where the relaxation time of the system increases beyond the threshold value of $\tau/\tau_0=10^{6}$. From Eq.\,(\ref{relation_time}) we see that $T_g$ gets modified due to activity as $T_g=E/(6\ln10)+T_K-Hf_0^2\tau_p/(1+G\tau_p)$, similar to $T_K^{eff}$.
Using this definition we compare our theory with the molecular dynamics simulation data~(\cite{mandal2016}, {\it Methods}), by rewriting Eq.\,(\ref{relation_time}) as
\begin{equation}
 \ln\left(\f{\tau}{\tau_0}\right)=\f{E}{(T-T_K)+\f{f_0^2}{\Lambda}} \label{relt_comp}
\end{equation}
where $\Lambda=H\tau_p/(1+G\tau_p)$ is a constant since $\tau_p$ is kept fixed. 
Analysis of simulation data gives $\tau_0=0.135$ and $T_K=0.28$. 
We obtain the values of the other two constant parameters by fitting our analytical expression to one particular data set, at $f_0=1.50$ (fitting to the data for other values of $f_0$ works equally well), and obtain $E=1.55$ and $\Lambda=9.8$. The Angell plot~\cite{angell1991} shown in Fig.\,\ref{reltime}(c), demonstrates the excellent agreement between theory and simulation data.
We emphasise here that the theoretical lines are {\em not} individual fits to the simulation data, since all the plots of Eq.\,(\ref{relt_comp}) use the same constant parameter values obtained from one initial fit. 
We obtain the fragility parameter $m\equiv -T{\partial \ln(\tau/\tau_0)}/{\partial T}\big|_{T=T_g}$~\cite{xia2000}, which becomes
\begin{equation}\label{fragility}
m=\f{(6\ln10)^2}{E}\left[\f{E}{6\ln10}+T_K-\f{Hf_0^2\tau_p}{1+G\tau_p}\right].
\end{equation}
In the simulations however,  the fragility parameter is obtained by fitting the data for the relaxation times with a form $\tau=\tau_0\exp[1/m(T/T_K^{eff}-1)]$, which differs from our expression [Eq.\,\ref{fragility})] by a constant. Using this form, we obtain 
\begin{equation}\label{fragility_sim}
 m=\f{T_K}{E}-\f{f_0^2}{\Lambda E}.
\end{equation}
Since it was not possible to obtain simulation data for very low $T$, there is some systematic error in the values of $m$ since the value of $m$ strongly depends on $\tau$ close to $T_g$, though we expect the functional dependence on activity to remain the same. We find that our theory underestimates the simulation data by a factor of 1.25, although the functional dependence on $f_0$ is predicted correctly, as shown in Fig.\,\ref{reltime}(d), where we have plotted Eq.\,(\ref{fragility_sim}) multiplied by a factor of $1.25$, using the same constants as in Fig.\,\ref{reltime}(c). 
As $f_0$ increases, the behaviour of $\tau$ becomes closer to Arrhenius law implying that the system becomes a stronger glass former~\cite{angell1991}.
Thus, if $\tau_p$ is fixed and activity is controlled through $f_0$, the system becomes a stronger glass former with increasing activity. 

\begin{figure}
\includegraphics[width=8.6cm]{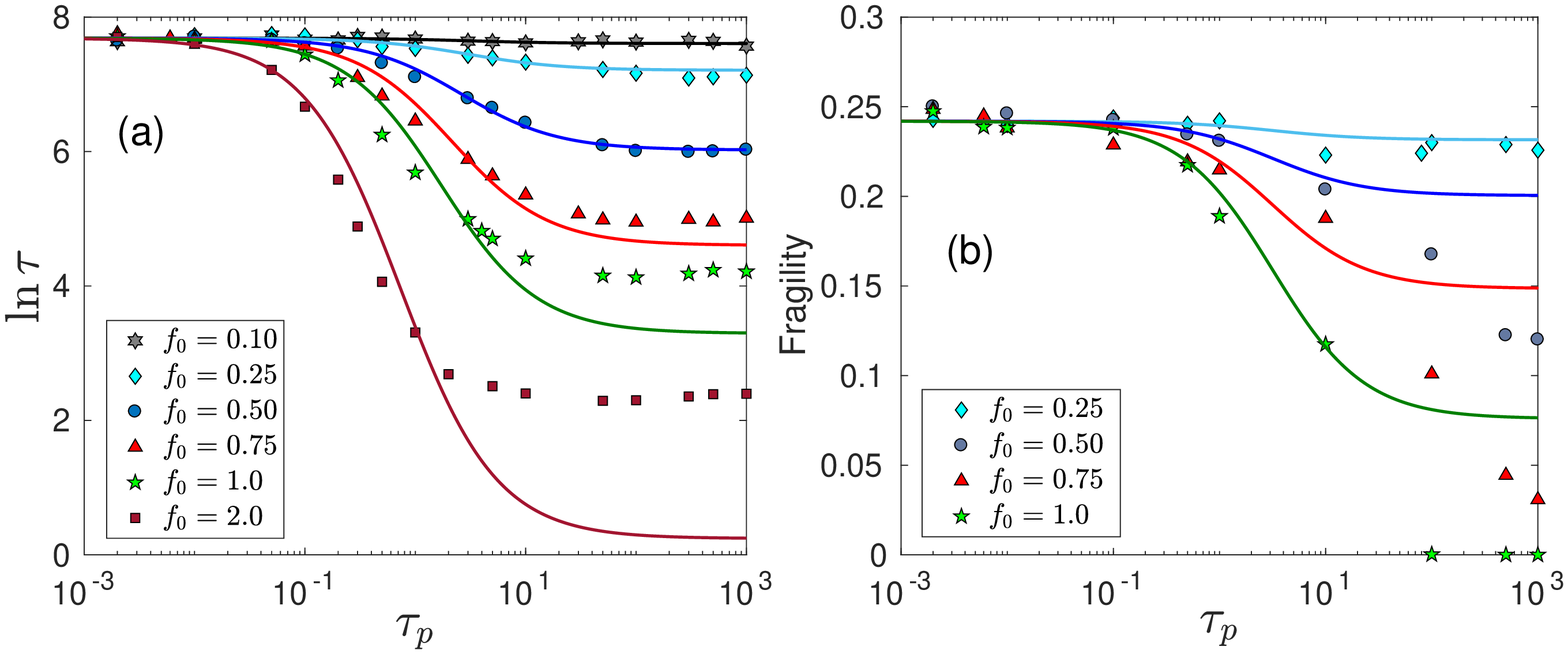}
\caption{Results for Model 1 when $\tau_p$ is the control parameter: (a) Relaxation time [Eq.\,(\ref{relation_time})] as a function of $\tau_p$ for different values of $f_0$. We test our theoretical predictions (lines) with the MD simulation results (symbols) 
using $E=1.55$, $T_K=0.28$, $H=0.042$ and $G=0.316$. The simulations were performed at $T=0.45$. The theory agrees well with simulation data at small $f_0$ and deviates at larger $f_0$ due to effects not included within the theory (see text).
(b) Fragility decreases with $\tau_p$ and saturates to a certain value. The parameters are the same as in Fig.\,3(a).}
\label{model1_results}
\end{figure}

\subsection*{Active RFOT confronts simulations - tuning $\tau_p$}
We next 
fix $f_0$ (or $T_{eff}^{sp}$) and control activity by changing $\tau_p$. Our active RFOT analysis of 
Model 1, where $\l$ and $\tau$ are given by Eqs.\,(\ref{length})\,and\,(\ref{relation_time}) respectively and the fragility is $m=T_K/E-Hf_0^2\tau_p/(1+G\tau_p)$, predicts that the relaxation time decreases for increasing $\tau_p$, while the dynamics become independent of $\tau_p$, when $\tau_p\gg 1/G$. 
We have tested these theoretical predictions by obtaining $\tau$ as function of $\tau_p$ from our simulations,
at different values of $f_0$. 
Figure\,\ref{model1_results}(a) shows this comparison using the values $H=0.0419$ and $G=0.3155$ - the corresponding value of 
$\Lambda$ that we obtain is approximately 1.25 times larger than the value used in the previous section (Fig.\,\ref{reltime}). We attribute this difference to the fact that in the previous section,
$T_g$ was obtained using a fit that is sensitive to the low-temperature data - this would be a less accurate estimate due to the long simulation time required.
The theoretical predictions agree very well with the simulations for low $f_0$ and $\tau_p$, and become systematically worse at larger activity, for example, at large $\tau_p$ for $f_0=1.0$ and $2.0$. 

Moreover, the effective medium approach to obtain the active contribution, Eq.\,(\ref{faeq2}), is appropriate for low $f_0$ and $\tau_p$~\cite{supmat}, and assumes that the effective friction coefficient ($\gamma$) and confining potential ($k$) are independent of the activity. This assumption would be expected to break down at larger activity, where the active particle finds it harder to move due to jamming effects. Our model therefore overestimates the value of the active correction to the configurational entropy compared to the simulations, and therefore underestimates the relaxation times, as we see in 
Fig.\,\ref{model1_results}(a).

In terms of an effective temperature $T_{eff}$, as defined in~\cite{supmat}, we expect a quadratic increase of $T_{eff}$ with $f_0$. It was found in~\cite{preisler2016} that for large $f_0$, $T_{eff}$  grows more slowly than the predicted quadratic increase (Fig.\,S3)~\cite{supmat}, and this deviation was attributed to activity-induced phase-separation.
We leave it for a future study to explore the properties of activity-driven jamming in dense assemblies of active particles. 

We can also compare the fragility parameter evaluated in the theory and simulations (Fig.\,\ref{model1_results}b). We again see fair agreement for small $f_0$ and $\tau_p$, and a clear deviation between theory and simulation data for larger $f_0$ and larger $\tau_p$, as discussed above.
Within this model, both $\tau$ and $m$ decrease as $\tau_p$ increases. Thus, $\tau_p$ fluidises the system and makes it a stronger glass former, similar to varying $f_0$.

\begin{figure}
 \includegraphics[width=8.6cm]{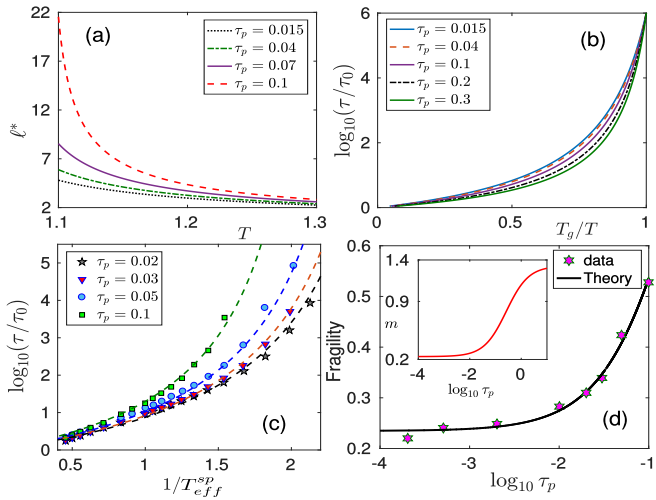}
 \caption{Results for Model 2 when $\tau_p$ is the control parameter. (a) $\l$ as a function of $T$ for different $\tau_p$ with $T_{eff}^{sp}=0.01$, using Eq.\,(\ref{lmodel2}).
 (b) We define $T_g$ when $\tau/\tau_0$ becomes $10^6$ and plot $\log(\tau/\tau_0)$ as a function of $T_g/T$ for $T_{eff}^{sp}=0.64$, using Eq.\,(\ref{taumodel2}). As $\tau_p$ increases, the curves depart further from the Arrhenius behaviour, which means the system becomes more fragile with increasing $\tau_p$.
(c) Relaxation time of the system as a function of $1/T_{eff}^{sp}$ (see text) at different $\tau_p$. We compare our theoretical results for $\tau$, with the simulation data of~\cite{flenner2016}. As the system dynamics in~\cite{flenner2016} comes from activity alone, $\tau_0$ becomes a function of $\tau_p$ when $\tau_p$ dominates the dynamics and we obtain $\tau_0=0.0615$, $0.0807$, $0.0992$ and $0.1863$ for $\tau_p=0.02$, $0.03$, $0.05$ and $0.1$ respectively~\cite{supmat}). (d) We expect the fragility to increase roughly linearly with $\tau_p$ from our theory (see text), and this matches quite well with the simulation data of \cite{flenner2016}. {\bf Inset:} Our theory predicts that the fragility saturates at large $\tau_p$.}
 \label{comp_flenner_etal}
\end{figure}

Next we consider Model 2, where $\l$ and $\tau$ are given by Eqs.\,(\ref{lmodel2})\,and\,(\ref{taumodel2}).
We set a constant $T_{eff}^{sp}$ and show the behaviour of $\l$ as a function of $T$ as we increase $\tau_p$ (Fig.\,\ref{comp_flenner_etal}a). 
At any value of $T$, we see that $\l$ increases monotonically as we increase $\tau_p$. Thus, $\tau_p$ promotes glassy behaviour in the active system in the sense that larger $\tau_p$ drives the system 
more towards the glassy state, as found in~\cite{flenner2016}. To display the behaviour of $\tau$ as a function of $\tau_p$, we show the Angell plot~\cite{angell1991} (Fig.\,3b).
We find that the curves depart further from the Arrhenius behaviour as $\tau_p$ increases, which means fragility increases with $\tau_p$.
From our theory, we obtain that the effective Kauzmann temperature, $T_K^{eff}$, glass transition temperature, $T_g$, and fragility, $m$, all increase with $\tau_p$ when $T_{eff}^{sp}$ remains constant. 
For a quantitative comparison of our theory with simulation data of~\cite{flenner2016}, we rewrite Eq.\,(\ref{taumodel2}) as in~\cite{supmat} (in the athermal case): 
\begin{equation}
\ln{\tau}=\ln{\tau_0}+\f{{E}}{[-{T}_K+T_{eff}^{sp}/(1+G\tau_p)]},
\end{equation}
where we have set $H$ to unity.
Since the dynamics of the system in~\cite{flenner2016} is a result of activity alone, $\tau_0$ becomes a function of $\tau_p$, and we fit a value of $\tau_0$ for each value of $\tau_p$, as was done in~\cite{flenner2016} (see Fig.\,S4~\cite{supmat}). Note that $\tau_0$  determines only the large $T_{eff}^{sp}$ limit, while our model gives precise prediction for the behaviour in the limit of small $T_{eff}^{sp}$, approaching the glassy regime.
We compare our theoretical results to the data of~\cite{flenner2016} in Fig.\,\ref{comp_flenner_etal}(c) using: ${E}=1.255$, ${T}_K=0.305$ and $G=3.801$. We also obtained the data for fragility $m$ from Fig.\,11(b) of~\cite{flenner2016} and fit it with our theoretical prediction for this model: $m=a-b/(1+G\tau_p)$, with $a=1.328$ and $b=1.093$ (Fig.\,\ref{comp_flenner_etal}d). Our theory predicts that $m$ saturates at large $\tau_p$ (Fig.\,\ref{comp_flenner_etal}(d) inset).
Irrespective of the detailed behaviour, the point is that this system becomes more {\it fragile} as $\tau_p$ increases. This is opposite to the behaviour of Model 1 (at small $\tau_p$). 
This is also different from the trend shown 
on increasing $T_{eff}^{sp}$ within this model.

\subsection*{Discussion}
We have developed an RFOT theory for active glasses, by proposing an analytic expression for the active contribution to the configurational entropy, using an effective medium approach that involves the dynamics of a typical 
active particle within a caging potential.
It will be interesting  to explore, in future simulations, the extent of validity of this single-particle approach and the relation between 
the effective parameters of this single-particle theory and the many-body properties of the system, such as the density and inter-particle correlations.
While our effective-medium, single-particle model captures  the dependence of the mean relaxation time on the activity parameters rather well, it shows deviations at large activities where activity-induced phase-separation and jamming occur. 
Our theory should be viewed as a first step towards a more rigorous analytical treatment of active glasses.

While we have approximated the effects of activity by a change in the potential energy, it
is also possible to define an effective temperature $T_{eff}$ that describes, in an approximate way, 
the effects of activity. For example, Eq.\,(\ref{relation_time}) can be interpreted as changing the 
temperature $T$ to $T_{eff} = T+(Hf_0^2\tau_p)/(1+G\tau_p)$, which implies that the effective
temperature increases quadratically with $f_0$. This is similar to the results reported in~\cite{preisler2016} which defined an effective temperature from a calculation of the
entropy of the active system. Both here (Fig. \ref{model1_results}) and in~\cite{preisler2016}, there are deviations from the model at large active forces~\cite{supmat}.


What are the implications of our results for cell or tissues that show glassy behaviour? 
To address this, we need to know the statistical nature of the active fluctuations in these systems.
In cells, activity is often realised in the form of active stresses
generated by ATP-driven motor proteins that bind/unbind onto an isotropic cytoskeleton (chemical energy transduced per ATP molecule under physiological conditions is $\approx 20$k$_B$T). 
 The effective temperature estimated from fluctuations \cite{fodor2015,fodor2016a} in the active stress arising from independent motors is proportional to $f_0^2 \tau_m$, where $\tau_m$ is the activity correlation time of the
 force generators~\cite{basu2008}. 
  This is precisely of the form Eq.\,(\ref{relation_time})
 obtained from Model 1,  following which we would expect that higher ATP decreases the fragility, driving the system towards a strong glass former, and that  activity, controlled through 
either $f_0$ or $\tau_p$, would always fluidise the system. 
This prediction agrees with early experiments~\cite{zhou2009,sadati2014} and
confirmed in a recent microrheology study~\cite{nishizawa2017} of the cytoplasm of a variety of cells, that shows strong glassy behaviour in wild type, which becomes
more fragile upon ATP  depletion. 
%

Beyond biology, active glasses present a challenge to physics theory, with the main advances in this field relying so far on numerical simulations. Refs.\,\cite{mandal2016,flenner2016} have produced data indicating the opposite roles of the active parameters.
Our model provides analytic expressions that  give excellent description of available numerical simulation results, and resolves the apparent contradiction between the effects of $f_0$ (or $T_{eff}^{sp}$) and $\tau_p$.
Since our theory is based on an effective medium approach for the dynamics of a caged active particle, we believe our qualitative results are likely to be independent of the precise form of the inter-particle interactions.
Our theory could then be made to confront precision experiments 
in synthetic systems~\cite{behchinger2016}, where the active forces and persistence time can be
controlled independently.

\subsection{Acknowledgements}
We thank S. Ramaswamy, P. G. Wolynes and S. Sastry for useful discussions. SKN thanks Koshland foundations for funding through a senior postdoctoral fellowship. RM acknowledge CSIR for financial help and Simons Centre at NCBS for computational resources and hospitality.


%

\section{Supplementary Material}

\renewcommand{\theequation}{S\arabic{equation}}
\renewcommand{\thefigure}{S\arabic{figure}}

\setcounter{equation}{0}
\setcounter{figure}{0}

Systems of self-propelled particles, 
associated with a self-propulsion force, $f_0$, and a persistence time, $\tau_p$,
are good models for a large class of active systems, that include living systems, such as a tissue comprising motile cells \cite{jacques2015SM,garcia2015SM}, or an ensemble of ants, fishes or birds \cite{sriramreviewSM,gravish2015SM} and nonliving systems, such as vertically vibrated polar rods \cite{nitin2014SM,dauchot2005SM}. 
The internal medium of the cell is associated with large effective viscosities owing to molecular overcrowding and hence can exhibit glassy behaviour. Similarly in the context of tissues, the high density
of cells can lead to jamming, a quintessential glassy phenomenon. Both these systems undergo a transition to fluidisation initiated by activity, and hence it is important to understand the glassy dynamics of such active systems.
In particular, we would like to understand how active {\it fluctuations} affect the approach to glassy behaviour in dense assemblies of active self-propelled particles.
One of the outcomes of our theoretical study is to highlight the fact that the detailed nature of activity is important in making predictions about the dependence of
activity on features of glassiness.

\subsection{Challenges in extending Random First Order Transition (RFOT) Theory to an active system}
Introducing activity within the basic RFOT picture~\cite{kirkpatrick1989SM} presents a number of difficulties:
\begin{itemize}
 \item 
 Active systems are inevitably out of equilibrium and it is not clear how far an extension of a theory that is based on equilibrium thermodynamics is going to be valid.
 \item
 As the term `active system' describes a diverse class of systems, where activity plays significantly different roles in different systems, a unified description is unrealistic at present.
 \item
 Configurational entropy density, $s_c(T)$, and the surface mismatch energy play the major roles within RFOT phenomenology and the precise nature how activity affects these quantities is not known.
 However, see \cite{preisler2016SM} for results on the entropy of active systems in the dilute limit, i.e., away from the glassy regime.
 \item
 It is not clear if the basic mosaic picture will survive under activity.
\end{itemize}

\begin{figure}
 \includegraphics[width=8.6cm]{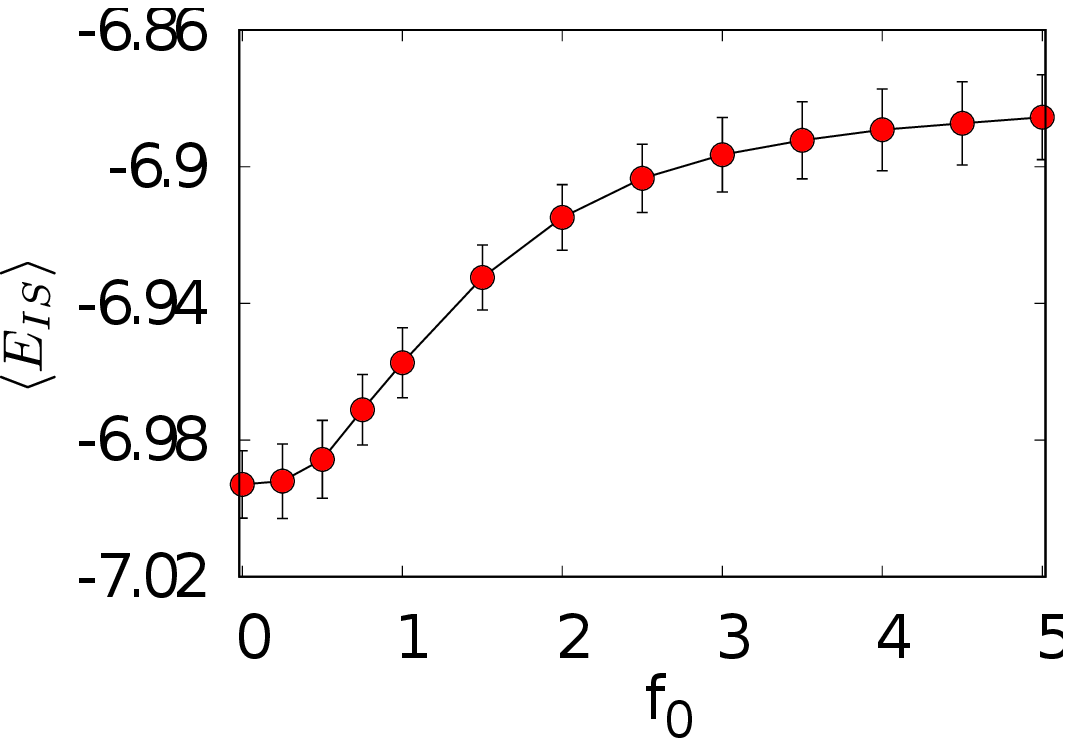}
 \caption{Average energy $\langle E_{IS}\rangle$ of the inherent structures as a function of activity strength $f_0$ at a fixed temperature for the model of Ref. \cite{mandal2016} show $\langle E_{IS}\rangle$ increases with increasing $f_0$. This implies that the configurational entropy also increases with $f_0$. The simulation was carried out for $N=1000$ particles with persistence time $\tau_p=4.0$ and fraction of active B particles $\rho_a=1$.}
 \label{EIS}
\end{figure}

Here we have extended the RFOT theory of conventional glasses to include the effects of activity in a minimal way. Since active systems 
are manifestly out of equilibrium, we must first find a physically reasonable definition of the configurational entropy $s_c$ for such systems. The configurational entropy of passive thermal systems is usually defined in terms of the multiplicity of the local minima of the potential energy (``inherent structures'') whose basins of interaction are visited by the system at a particular temperature $T$. Here, we continue to use the same definition of the configurational entropy for active systems. Particle configurations in a steady state of an active systems can be used as starting points of a minimisation procedure to locate a set of inherent structures characteristic to the steady state being considered. The multiplicity of inherent structures obtained this way would be different from that of the inherent structures obtained for an identical system without activity at the same temperature. Therefore, the temperature dependence of the configurational  entropy of an active system would be different from that for its passive counterpart. 
We have obtained the average energy, $\langle E_{IS}\rangle$, of inherent structures as a function of activity strength $f_0$ for the model of Ref. \cite{mandal2016SM} and find that $\langle E_{IS}\rangle$ increases with $f_0$ at a fixed temperature as shown in Fig. \ref{EIS}.
In the next section, we present the details of how we compute the activity corrections to the configurational entropy within a specific model of a system of self-propelled particles.

\subsection{Active contribution to configurational entropy}
Following~\cite{xiawolynesSM,sastrySM,berthiercoslovicSM} and the discussion above, we take the configurational entropy density $s_c[\Phi]$ 
 to depend on the many-body potential $\Phi$ at temperature $T$.

We start with the overdamped dynamics of the system of passive particles, given by,
 \be
 \partial_t {\bf x}_i = - \mu \nabla_i \Phi_i + {\bf f}_i(t)
 \label{eq:passive}
 \ee
 where ${\bf x}_i$ is the position of the $i$-th particle, $\Phi_i \equiv \sum_{j\neq i} \phi({\bf x}_i, {\bf x}_j)$ is the many-body potential experienced by particle $i$ and ${\bf f}_i(t)$ is the thermal noise
  (gaussian, white noise with zero mean).  Defining the full many-body potential as ${\Phi} = \f{1}{2}\sum_i {\Phi}_i$, 
  Refs.\,\cite{xiawolynesSM,sastrySM,berthiercoslovicSM,senguptaSM} would suggest that  the configuration entropy depends on the steady state profile of this
  $\Phi$ : $s_c=s_c[\Phi, T]$ (where $\Phi$ is understood to be averaged over the steady state distribution corresponding to Eq.\,(\ref{eq:passive})).
At this stage, we need not know how to compute this; all we need is that this is somehow given.
With this in hand, we discuss an effective medium approach that goes into the calculation of the activity corrected configuration entropy density $s_c[{\tilde \Phi}]$ 
needed for the active version of the RFOT theory described in the main text.

 To this end, we first note that  the 
  corresponding dynamical equation for active self-propelling particles is,
    \be
 \partial_t {\bf x}_i = - \mu \nabla_i \Phi_i + {\bf f}^{a}_i(t)+ {\bf f}_i(t)
  \label{eq:actve}
 \ee
 where ${\bf f}^{a}(t)$ is an active noise taken to be exponentially correlated in time, with zero mean. One can interpret the above equation~\cite{fodor2016SM,marconi2016SM} as modifying the 
  effective many-body potential experienced by particle $i$ by,
  \be
  {\tilde \Phi}_i = \Phi_i + \delta \Phi_i \equiv  \Phi_i - \frac{1}{\mu} {\bf f}^{a}_i \cdot {\bf x}_i\,.
  \label{eq:effpot}
  \ee
  
    The corresponding configurational entropy density for the active system is
 $s_c({\tilde \Phi})$, which we expand in a Volterra series (or a functional Taylor series) as,
\bea
 s_c[{\tilde \Phi}] &=  & s_c[\Phi] +  \int  \delta \Phi(\{{\bf x}_i\}) \frac{\delta s_c[\Phi]}{\delta \Phi(\{{\bf x}_i\})}\bigg|_{{\bf f}^{a}=0,T}\,\prod_i d{\bf x}_i \nonumber \\ 
& &  + \ldots 
\label{eq:volterra}
 \eea
 In what follows, we will retain only the first term, assuming $f^{a}$ is small. However, evaluating this first-order correction, even approximately, poses a problem.
\\

\noindent
{\it Effective-medium approach} :   To proceed we first
approximate the above many-body dynamics by the dynamics of a typical self-propelled particle
with coordinate ${\bf x}$ moving
 in an effective caging
potential $V(\sigma)$, where the collective
 variable $\sigma$, is assumed to evolve much slower than ${\bf x}$, i.e., over a time scale equal to or greater than the $\alpha$-relaxation time, $\tau_{\alpha}$.
Here we take the caging potential to be harmonic and static, the overdamped Langevin dynamics is then given by 
   \be
 \partial_t {\bf x} = - {\gamma}^{-1} \nabla V(\sigma) + {\gamma}^{-1}{\bf f}^{a}(t)
 \label{eq:singlepart}
 \ee
 where ${\gamma}$ is the effective particle friction coefficient.  This assumes that $\tau_p \ll \tau_{\alpha}$.
 
We will now approximate the first-order correction in Eq.\,(\ref{eq:volterra}) by $\kappa_a \,\langle \delta \Phi\rangle$, where 
$\kappa_{a} \equiv \bigg\langle {\frac{\delta s_c}{\delta \Phi}}\bigg|_{{\bf f}^{a}=0,T}\bigg\rangle$, an {\it active fragility parameter}, and $\langle \ldots \rangle$ is an average over the steady state distribution corresponding to Eq.\,(\ref{eq:singlepart}).   This parameter measures the sensitivity of the configurational entropy 
to changes in the potential or alternatively, the active force. For simplicity, we will take it to be a constant fit parameter (although there is no apriori justification for this).
The $\langle \delta \Phi\rangle$ term can be easily evaluated
by solving the corresponding Fokker-Planck equation, which we will now do for the two models of active forcing described in the main text.

The motivation for identifying the active correction to the configurational entropy with the change in potential energy of a trapped active particle, can be as follows \cite{noneqmctSM}: due to activity the particle explores the space around it to a larger extent, and this can be quantified by the change in potential energy, similar to an ``effective temperature'' (see below).

\noindent
{\bf Model 1:}  \,\, $\langle {\bf f}^a_i(0) \cdot {\bf f}^a_i(t)\rangle = f_0^2\exp[-t/\tau_p]$, where $f_0$ is the average strength and $\tau_p$ is the persistence of the active noise.

With this noise statistics, Eq.\,(\ref{eq:singlepart})
 written for a harmonic caging potential in one dimension, is equivalent to the following Ornstein-Uhlenbeck process \cite{szamel2014SM, mandal2016SM},
\bea
\dot{x}(t)& = & -\frac{k}{\gamma} x(t)+ \frac{f_0}{\gamma} \psi(t) \\
\dot{\psi}(t) & = & -\frac{\psi(t)}{\tau_p}+ \frac{1}{\sqrt{\tau_p}} \eta(t)
\label{eq:oup1}
\eea
 where $k$ is the spring constant of the harmonic potential,
 and the noise $\eta$ has zero mean and is delta-correlated and white : $\langle \eta(t) \rangle=0$ and $\langle \eta(t) \eta(t^{\prime}) \rangle=2 \delta(t-t^{\prime})$. The calculation below is qualitatively identical to the case with fixed active force amplitude \cite{benisaac2015SM}.
 

The corresponding Fokker-Planck equation for the joint probability distribution, $P(x, \psi, t)$,  is \cite{gardinerSM},
\begin{equation}
\begin{split}
\frac{\partial P(x,\psi)}{\partial t}=\frac{c}{\tau_p} \frac{\partial^2 P(x,\psi)}{\partial \psi^2}+\(\frac{k}{\gamma}+\frac{1}{\tau_p}\) P(x,\psi) \\+  \(\frac{kx}{\gamma} -
\frac{f_0 \psi}{\gamma}\) \frac{\partial P}{\partial x}+ \frac{1}{\tau_p} \psi \frac{\partial P}{\partial \psi}
\end{split}
\end{equation}
whose steady state is easily seen to be \cite{gardinerSM},
\begin{equation}
P^{ss}(x,\psi)=A \exp \[ -c_1 x^2 -c_2 \psi^2 -c_3 x \psi \]
\end{equation}
with $c_1=\frac{k}{2 \gamma \tau_p  f_0^2}(k \tau_p +\gamma)^2$, $c_2=\frac{k \tau_p+\gamma}{2  \gamma}$ and $c_3=\frac{-k(k \tau_p +\gamma)}{f_0  \gamma}$, and
$A$ is the normalization factor.

This can be simply integrated over $\psi$, to obtain the marginal distribution,
\begin{equation}
P^{ss}(x)\propto \exp\[- \frac{kx^2}{2} \(\frac{k \tau_p +\gamma}{f_0^2 \tau_p}\)\]
\end{equation}
which has a Boltzmann form, $P^{ss}(x) \propto \exp\[ -\frac{k x^2}{2 T_{eff,a}} \]$, with an effective temperature \cite{benisaac2015SM}, 
\be\label{activeteffmod1}
T_{eff,a}=\frac{f_0^2 \tau_p/\gamma}{1+ k \tau_p/\gamma}\,.
\ee
 
 Using the steady state joint probability distribution $P(x,\psi)$, we can show that for Model 1,
 \begin{equation}
\langle \psi x \rangle= \frac{k}{f_0} \langle x^2 \rangle \equiv  \frac{T_{eff,a}}{f_0}\, .
\end{equation}



\noindent
{\bf Model 2:}  \,\, $ \langle {\bf f}_a(0) \cdot {\bf f}_a(t)\rangle = (T_{eff}^{sp}/\tau_p)\exp[-t/\tau_p]$, where $\tau_p$ is the persistence of the active noise.

With this noise statistics, Eq.\,(\ref{eq:singlepart})
 for a harmonic caging potential, is equivalent to the following Ornstein-Uhlenbeck process \cite{szamel2014SM, mandal2016SM},
\bea
\dot{x}(t) & = & -\frac{k}{\gamma} x(t)+ \frac{1}{\gamma} \psi(t) \\
\dot{\psi}(t) & = & -\frac{\psi(t)}{\tau_p}+\frac{\sqrt{{T}^{sp}_{eff}}}{\tau_p} \eta(t)
\label{eq:oup2}
\eea
where the noise $\eta$ has zero mean and is delta-correlated, white :   $\langle \eta(t) \rangle=0$ and $\langle \eta(t) \eta(t^{\prime}) \rangle=2 \delta(t-t^{\prime})$. 


The corresponding Fokker-Planck equation for the joint probability distribution, $P(x, \psi, t)$,  is \cite{gardinerSM},
\begin{equation}
\begin{split}
\frac{\partial P(x,\psi)}{\partial t}=\frac{{T}^{sp}_{eff}}{{\tau_p}^2} \frac{\partial^2 P(x,\psi)}{\partial \psi^2}+\(\frac{k}{\gamma}+\frac{1}{\tau_p}\) P(x,\psi) \\+  \(\frac{kx}{\gamma} -
\frac{\psi}{\gamma}\) \frac{\partial P}{\partial x}+ \frac{1}{\tau_p} \psi \frac{\partial P}{\partial \psi}
\end{split}
\end{equation}
whose steady state solution is,
\begin{equation}
P^{ss}(x,\psi)=A \exp \[ -c_1 x^2 -c_2 \psi^2 -c_3 x \psi \]
\end{equation}
with  $c_1=\frac{k}{2 \gamma {T}^{sp}_{eff}}(k \tau_p +\gamma)^2$, $c_2=\frac{(k \tau_p+\gamma)\tau_p}{2 \gamma {T}^{sp}_{eff}}$ and $c_3=\frac{-k(k \tau_p +\gamma)\tau_p}{\gamma {T}^{sp}_{eff}}$.
The marginal distribution is, 
\begin{equation}
P^{ss}(x)\propto \exp\[- \frac{kx^2}{2} \(\frac{k \tau_p +\gamma}{{T}^{sp}_{eff}}\)\]
\end{equation}
which again has a Boltzmann form, $P^{ss}(x) \propto \exp\[ -\frac{k x^2}{2 T_{eff,a}} \]$ with an effective temperature 
\be\label{activeteffmod2}
T_{eff,a}=\frac{T^{sp}_{eff}/\gamma}{1+k \tau_p/\gamma}\,.
\ee
 
 Using the steady state joint probability distribution $P(x,\psi)$, we can show that for Model 2,
 \begin{equation} \label{model2_activesc}
\langle \psi x \rangle= k \langle x^2 \rangle \equiv T_{eff,a} \, .
\end{equation}

\subsection{Comparison with recent simulation in terms of effective temperature ($T_{eff}$)}

\begin{figure}
 \includegraphics[width=8.6cm]{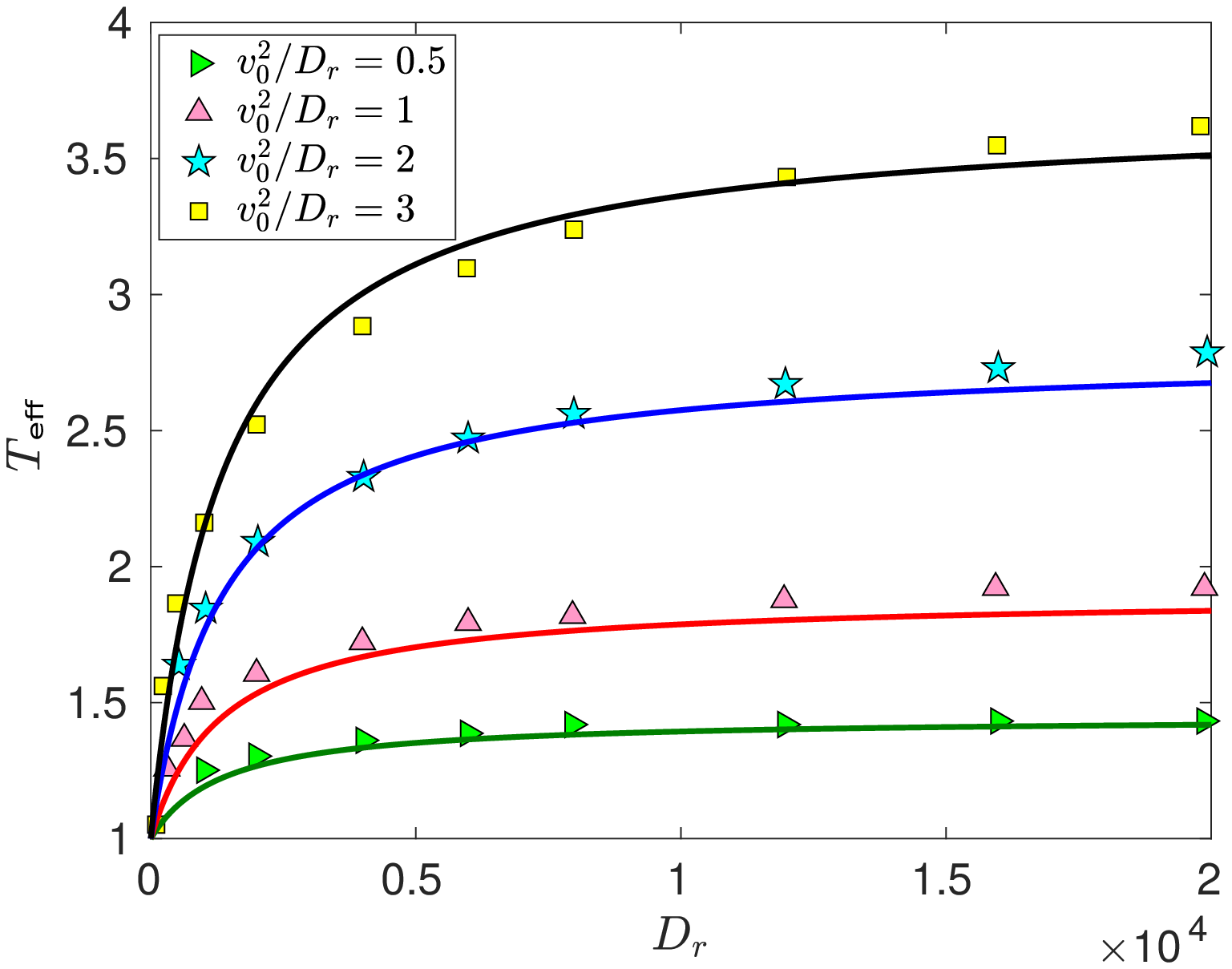}
 \caption{Our effective medium approach gives rise to an effective temperature (Eq.\,(\ref{activere2})) that we compare with simulation results of~\cite{preisler2016SM}. Symbols are data obtained from Fig.\,4(b) of~\cite{preisler2016SM} and solid lines are plots of Eq.\,(\ref{activere2}) with $A=0.893951$ and $B=1350.5$.}
 \label{comp_PD}
\end{figure}

Within  the effective medium approach outlined above, the steady state distribution always has a Boltzmann form, and hence we can assign an effective temperature, as in
Eqs.\,(\ref{activeteffmod1})\,and\,(\ref{activeteffmod2}). Such effective temperature descriptions have been discussed in~\cite{loi2011SM,szamel2014SM,wang2011SM,benisaac2011SM,shen2004SM,preisler2016SM,redner2013SM,redner2016SM}.
This is fortuitous and may not always be possible~\cite{fodor2016SM}, especially in active dense many-body systems.

We therefore compare the effective temperature definition in a many-body active system~\cite{preisler2016SM}, with the effective medium approach. Ref. \cite{preisler2016SM} has proposed a method to calculate the entropy of an active system consisting of Brownian spheres and show that active systems can be mapped to equilibrium systems with a many-particle distribution function akin to Boltzmann distribution, but characterised by an effective temperature.
%
Ref.\,\cite{preisler2016SM} considers the coefficient of rotational Brownian motion, $D_r$, as a measure of persistence, $D_r\sim 1/\tau_p$. Using Eq.\,(\ref{activeteffmod1}) we obtain $T_{eff}$ in terms of $D_r$ and $v_0$, the self-propulsion velocity, as
\begin{equation}\label{activere2}
 T_{eff}=T+T_{eff,a}=T+\f{A(v_0^2/D_r) D_r}{B+D_r}
\end{equation}
where $T$ is the equilibrium temperature (set to unity) and $A$ and $B$ are two constants. We have included the factor $2D_t$, as defined in \cite{preisler2016SM}, in $A$ above since $D_t$ remained constant in the simulation. We obtain the constant parameters $A=0.893951$ and $B=1350.5$ by fitting Eq.\,(\ref{activere2}) with the data corresponding to $v_0^2/D_r=0.5$ in Fig.\,4(b) in~\cite{preisler2016SM}
and show the behavior of Eq.\,(\ref{activere2}) along with the data of~\cite{preisler2016SM} in Fig.\,\ref{comp_PD}. Note that the theoretical analysis in~\cite{preisler2016SM} was done in the limit $\tau_p\to 0$. This comparison shows the consistency of our approximation.

\begin{figure}
 \includegraphics[width=8.6cm]{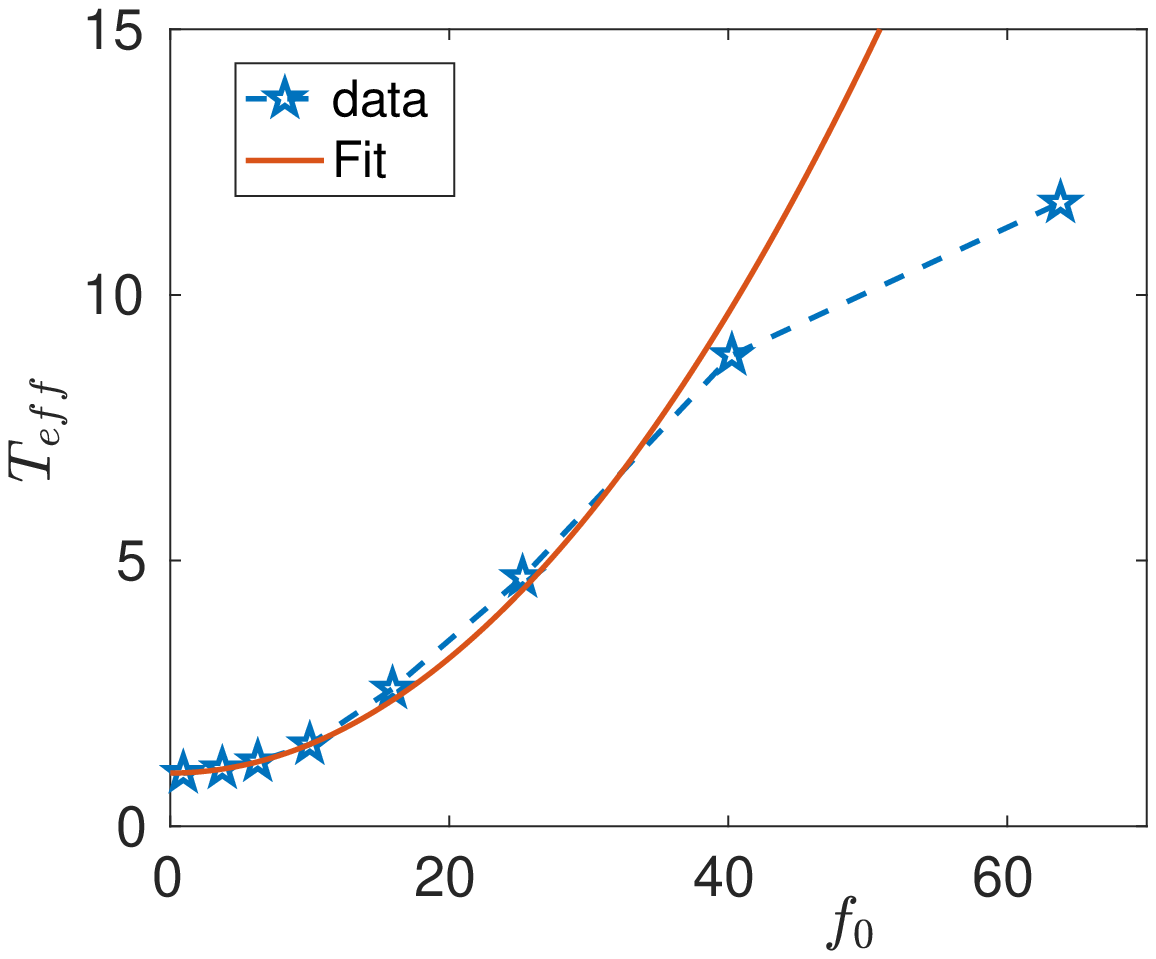}
 \caption{Trend of $T_{eff}$ as a function of $f_0$ in simulation (data taken from Fig.\,5(b) of~\cite{preisler2016SM} N=1024). Our theoretical result (Eq.\,\ref{activere2}), that $T_{eff}$ behaves quadratically with $f_0$ breaks down at large $f_0$. It is possibly because activity induced phase separation is important at large activity and this mechanism is absent within our theory.}
 \label{effectiveT_dev}
\end{figure}

We have found that our theory deviates from the simulation results at large activity (Fig. 3 in the paper). When activity becomes large, the system is driven far away from equilibrium and our
effective medium approach breaks down.
 Ref.\,\cite{preisler2016} also finds such a deviation; we have plotted in Fig. \ref{effectiveT_dev} the data from Fig.\,5(b) of~\cite{preisler2016SM}, and the fit with our model $T_{eff}\sim 1+a f_0^2$ (Eq. \ref{activere2}) with $a=0.0054$ being a fitting parameter for small $f_0$.

\subsection*{Simulation model}
We follow the simulation strategy as detailed in~\cite{mandal2016}.
We perform a molecular-dynamics (MD) simulation of a Kob-Andersen binary mixture~\cite{kob1994} of A and B-type particles in the ratio (80:20) interacting via a Lennard-Jones pair potential,
\begin{equation}
 V_{ij}(r)=4 \epsilon_{ij} \left[\left(\frac{\sigma_{ij}}{r}\right)^{12}-\left(\frac{\sigma_{ij}}{r}\right)^{6}\right],
\label{eq:potential}
\end{equation}
where $r$ is the distance between two particles and indices $i,j \in \{A,B\}$.
The units of length and energy are determined by setting $\sigma_{AA}=1$ and $\epsilon_{AA}=1$. 
The rest of the parameter values  are taken to be:  $\sigma_{AB}=0.8 \sigma_{AA}$, $\sigma_{BB}=0.88 \sigma_{AA}$, $\epsilon_{AB}=1.5  \epsilon_{AA}$, $\epsilon_{BB}=0.5 \epsilon_{AA}$. The mass of the particles are $m_A=m_B=1.0$ and the overall number density of the system is $\rho=1.2$. The interaction potential is cutoff at $r_{ij}^c=2.5 \sigma_{ij}$ and it is smoothed with a quadratic function so that both the energy and forces are continuous at the $r_{ij}^c$. The particles follow Newton's laws; the equations of motion are integrated by employing the velocity version of the Verlet algorithm.

A fraction $\rho_a$ of B-type particles ($0\leq \rho_a \leq 1$) are subject to a propulsion force, whilst keeping the rest of the particles passive (here, we set $\rho_a=1$). 
Self-propulsion forces of the form ${\bf{f_0}}=f_0(k_x\hat{{\bf x}}+k_y\hat{{\bf y}}+ k_z\hat{{\bf z}})$ are randomly assigned to the active B-particles, with $k_x, k_y, k_z$ chosen randomly to have values $\pm 1$, so
as to conserve the net momentum of the system. The active particles are driven in the directions of $\{\bf{f_0}\}$ for a persistence time $\tau_{p}$, the directions of  $\{\bf{f_0}\}$ are then randomised by choosing a different set of $k_x, k_y, k_z$. We have checked that the dynamics driven by this 8-state clock realisation of the random propulsion forces, has the same qualitative features as a continuous O(3) realisation. Simulations here are for  $N=1000$ particles; we have checked that finite size effects are negligible.

\subsection*{Dynamical quantitites}
The slow dynamics is captured by  the two-point correlation function, $Q(t)$, defined as:
\begin{equation}
Q(t) = \frac{1}{N} \sum_{i}\langle w(\vert{\bf r}_i(t_0)- {\bf r}_i(t+t_0)\vert)\rangle
\end{equation}
where,
\begin{equation}
\nonumber
w(r)=
\left\{
        \begin{array}{ll}
                1  & \mbox{if } r \leq a\\
                0  & \mbox{otherwise}.
        \end{array}
\right.
\end{equation}
Here the $\langle \ldots \rangle$ is an average over number of particles $N$ and time origin $t_0$. The parameter $a$ is associated with the typical amplitude of vibrational motion of the particles. We have used $a=0.3$ for our analyses. The decay of $Q(t)$ in time is a measure of the dynamical slowing down and defines the $\alpha$-relaxation time $\tau_\alpha$, via $Q(\tau_{\alpha})=1/e$. The $\alpha$-relaxation time at different temperatures for a given $f_0$ can be fitted  to the Vogel-Fulcher-Tamman (VFT) form,
\begin{equation}
\tau_\alpha=\tau_\infty \exp \[\frac{1}{\kappa\(\frac{T}{T_{\mbox{\tiny{VFT}}}}-1\)} \]
\label{eq:VFT}
\end{equation}
where $\tau_\infty$ is the relaxation time at high temperatures and $\kappa$ is the kinetic fragility. The relaxation time extrapolates to infinity at $T_{\mbox{\tiny{VFT}}}(f_0)$, the 
putative glass transition temperature.

\subsection{Details for the results}
Within Model 1, we have the equations governing the most probable mosaic length scale, $\l$, and relaxation time $\tau$ as
\begin{align}
& \l=\left[\f{D}{(T-T_K)+\f{Hf_0^2\tau_p}{1+G\tau_p}}\right]^{2/d} \label{lengthSM}\\
& \ln\left[\f{\tau}{\tau_0}\right]=\f{E}{(T-T_K)+\f{Hf_0^2\tau_p}{1+G\tau_p}}. \label{relation_timeSM}
\end{align}
To compare with simulation data, we obtain the data from Fig.\,8 of Ref.\,\cite{mandal2016SM} and write Eq.\,(\ref{relation_timeSM}) as
\begin{equation}
 \ln({\tau}/{\tau_0})={E}/[(T-T_K)+\f{f_0^2}{\Lambda}] \label{relt_compSM}
\end{equation}
and obtain different constants through fitting the above equation with a particular data corresponding to $f_0=1.5$. In principle, we could also obtain these parameters from the detailed microscopic knowledge of the system that we didn't attempt here.
The excellent agreement between the theoretical predictions and simulation data shows that our theory captures the basic physics of the system. 
In this work we have assumed $\bar{\gamma}$ to be same as $\gamma_{passive}$, the surface mismatch energy of a passive system, for simplicity.
We also tried different forms for $\gamma_A\equiv (\bar{\gamma}-\gamma_{passive})=f(f_0,\tau_p)$. When we considered the amplitude of $\gamma_A$ to be small, the detailed form of $f(f_0,\tau_p)$ didn't alter the results qualitatively which is expected from Eq.\,(4). We obtained the best fit with a vanishing $\gamma_A$ as shown in Figs.\,2(c)\,and\,(d). It is possible that $\gamma_A$ has a non-trivial dependence on activity, but it is sub-dominant for the model we considered such that it doesn't play a major role in the glassy properties of the system.

\begin{figure}
 \includegraphics[width=8.6cm]{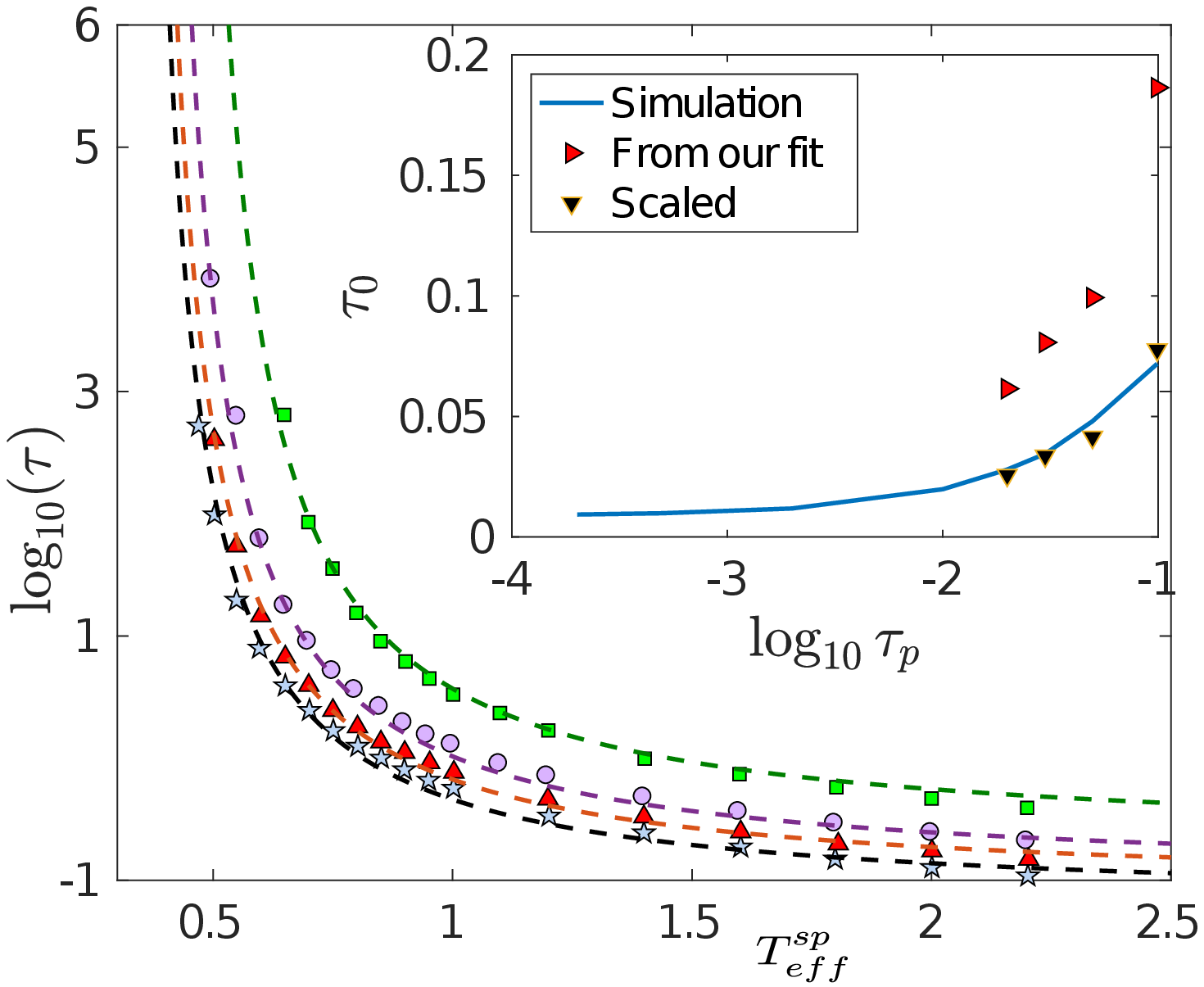}
 \caption{Relaxation time $\tau$ as a function of $T_{eff}^{sp}$. Data obtained from Ref. \cite{flenner2016SM} and the dotted lines are plots of Eq. (\ref{comp_flennerSM}) neglecting the second-order term in $\tau_p$ in the denominator with ${E}=1.255$, ${T}_K=0.305$, $G=3.801$ and $\tau_0(\tau_p)=0.0615$, $0.0807$, $0.0992$ and $0.1863$ for $\tau_p=0.02$, $0.03$, $0.05$ ans $0.1$ respectively. 
 Symbols: $\Square: \tau_p=0.1$, $\Circle: \tau_p=0.05$, $\triangle: \tau_p=0.03$, $\openbigstar: \tau_p=0.02$. {\bf Inset:} Plots of $\tau_0$ as a function of $\tau_p$. Symbols are the values obtained by fitting the theoretical expression, Eq.\,(\ref{comp_flennerSM}), with the simulation data where the scaled values are $\tau_0/2.5$. Solid line corresponds to the data in Fig.\,9(a) in Ref.\,\cite{flenner2016SM}.}
 \label{comp_flennerdata}
\end{figure}

For a quantitative comparison with simulation data when activity is controlled through $\tau_p$, we compare our theory with Ref.\,\cite{flenner2016SM}, which considers an athermal system of self-propelled particles where the dynamics of the system solely comes from activity. The active force in the simulation of~\cite{flenner2016SM} obeys
\begin{equation}
 \tau_p\dot{f}(t)=-f(t)+\eta
\end{equation}
where $\eta$ is a Gaussian white noise with zero mean and variance $\langle\eta(t)\eta(t')\rangle=2\xi_0 T_{eff}^{sp}\delta(t-t')$ (See Eq.\,2 in Ref.\,\cite{flenner2016SM}). With this active force that belongs to Model 2 in our classification and therefore using Eq.\,(\ref{model2_activesc}) we obtain the relaxation time as
\begin{equation}
 \ln{\tau}=\ln{\tau_0}+\f{{E}}{[-{T}_K+T_{eff}^{sp}/(1+G\tau_p)]}\label{comp_flennerSM}
\end{equation}
where we have set $H$ to unity.
Since Ref.\,\cite{flenner2016SM} considers a system where temperature doesn't play any role, we have set $T=0$. 
We obtain the data for relaxation time from Fig.\,6 of Ref.\,\cite{flenner2016SM} for $\tau_p=0.02$, $0.03$, $0.05$ and $0.1$. 
As the dynamics solely arises due to activity, $\tau_0$ becomes a function of $\tau_p$ when $\tau_p$ dominates the dynamics. Since $\tau_p$ is not large, we neglect the second-order term in $\tau_p$ in the denominator of Eq.\,(\ref{comp_flennerSM}) and through fitting we obtain ${E}=1.255$, ${T}_K=0.305$, and $G=3.801$ and $\tau_0(\tau_p)=0.0615$, $0.0807$, $0.0992$ and $0.1863$ for $\tau_p=0.02$, $0.03$, $0.05$ ans $0.1$ respectively. These values of $\tau_0$ are approximately $2.5$ times higher than the value of $\tau_0$ obtained in \cite{flenner2016SM} through the fitting of a different equation and a different temperature regime, hence, 
it's not surprising that their absolute values are different, however, we find that they are proportional. In the inset of Fig.\,\ref{comp_flennerdata} we show $\tau_0$ as obtained in Ref.\,\cite{flenner2016SM} by the solid line, $\tau_0$ obtained by our fit by $\rhd$ and we plot these values scaled by 2.5 as shown by $\triangledown$.
We show the comparison of our theory with the simulation data in Fig.\,\ref{comp_flennerdata}. In the main paper, we chose to present the same plot as $\log(\tau/\tau_0)$ as a function of $1/T_{eff}^{sp}$ as we wanted to emphasise the dependence of $\tau$ on $\tau_p$ in the low-temperature regime.

\end{document}